\documentclass{emulateapj}
\usepackage{natbib}
\shorttitle{Building massive compact planetesimal disks}
\usepackage{color}
\usepackage{amsmath}

\begin{document}
\title{Building massive compact planetesimal disks from the accretion of pebbles}
\author{John Moriarty}
\affil{Department of Astronomy, Yale University, New Haven, CT 06511, USA}
\email{john.c.moriarty@yale.edu}
\and
\author{Debra Fischer}
\affil{Department of Astronomy, Yale University, New Haven, CT 06511, USA}

\begin{abstract}
We present a model in which planetesimal disks are built from the combination of planetesimal formation and accretion of radially drifting pebbles onto existing planetesimals. In this model, the rate of accretion of pebbles onto planetesimals quickly outpaces the rate of direct planetesimal formation in the inner disk. This allows for the formation of a high mass inner disk without the need for enhanced planetesimal formation or a massive protoplanetary disk. Our proposed mechanism for planetesimal disk growth does not require any special conditions to operate. Consequently, we expect that high mass planetesimal disks form naturally in nearly all systems. The extent of this growth is controlled by the total mass in pebbles that drifts through the inner disk. Anything that reduces the rate or duration of pebble delivery will correspondingly reduce the final mass of the planetesimal disk. Therefore, we expect that low mass stars (with less massive protoplanetary disks), low metallicity stars and stars with giant planets should all grow less massive planetesimal disks. The evolution of planetesimal disks into planetary systems remains a mystery. However, we argue that late stage planet formation models should begin with a massive disk. This reinforces the idea that massive and compact planetary systems could form in situ but does not exclude the possibility that significant migration occurs post-planet formation.

\end{abstract}

\section{Introduction}

One of the most interesting findings of the Kepler mission is the prevalence of super-Earth and Neptune sized planets on short period orbits. Many of these planets reside in compact multi-planet systems often referred to as STIPs (Systems with Tightly packed Inner Planets).  Although such planets are lacking in the solar system, their ubiquity in the Kepler data indicates they are a natural outcome of the planet formation process.

\citet{Chiang13} found that the minimum surface density of material needed to form many of the Kepler systems is at least 5-10 times higher than the surface density of the minimum mass solar nebula (MMSN) extrapolated inward of Mercury's orbit. This high concentration of mass within the inner 1 AU or so of these systems is a strong indication that the transport of solid material in protoplanetary disks is an important aspect of planet formation. The difficulty lies in determining how and when this material is transported.

One hypothesis is that planets form first and then migrate to their observed locations. In this scenario planets would form farther out in the disk and migrate inwards due to gravitational torques exerted on them by the disk \citep[see][for review]{Kley12}. N-body simulations of planet formation that include migration forces have successfully produced systems of, typically 2-5, planets on close in orbits \citep[e.g][]{Terquem07, Hands14}. However, the formation with migration scenario is not without its problems. Most notably, these simulations produce an over-abundance of neighboring planets that are at or near mean motion resonances when compared to the Kepler sample. It is possible that additional forces that act after the planets have reached their final position may remove planets from resonance \citep[e.g.][]{Lithwick12, Batygin13}. Another prediction of the migration scenario is that planets form beyond the ice line and consequently will be very water rich. Observations of planet masses and radii find that of the planets with masses known to better than 20\%, the smaller planets (those with radii less that 1.7 R$_{\oplus}$) have densities that are consistent with an Earth-like composition \citep{Dressing15}.

Another possibility, promoted by \citet{Hansen12}, is that solid material migrated through the disk before planet formation and formed a massive planetesimal disk. Planets then formed from this disk in a similar manner as we believe the terrestrial planets of the solar system did. \citet{Hansen12} simulated the in situ formation of close in super-Earths and Neptunes from a massive disk and found good agreement between the observed distribution of planets and their simulated systems both in terms of the mass-period distribution and more detailed statistics. The major downside to this hypothesis is the omission of the physics of migration, which is believed to be important for bodies larger than Earth.

One of the difficulties in determining which, if either, of these scenarios is correct is our lack of knowledge of the initial conditions of planet formation. Specifically, the surface density of solids throughout the disk directly preceding planet formation is poorly known. This uncertainty adds a huge amount of freedom to planet formation models which is essentially unbounded.

Our poor knowledge of the surface density distribution in planetesimal disks is understandable considering the lack of observational constraints. What information we do have comes from the remnants of the solar system's planetesimal disk. However, we can still make predictions of what we expect planetesimal disks to look like which can offer some constraint to planet formation models.

One crucial piece of information gleaned from meteorites is that planetesimal disks are built up over time - over the course of about 2-3 million years \citep{Amelin02}. With this knowledge, we can think of planetesimal disk formation as an evolutionary sequence and not just some scaling of the gas/dust content of a protoplanetary disk at some point in time. In this context, the growth of a planetesimal disk depends on the rates of planetesimal formation, growth, destruction and migration. Integrating these rates over time, throughout the disk, provides us with the surface density of planetesimals to use as initial conditions for planet formation models.

In order to understand the growth of the planetesimal disk, we need to know what the reservoir of material is that is fueling this growth. Models of dust evolution in protoplanetary disks predict rapid growth of grains to millimeter and centimeter-size pebbles. \citep[e.g.][]{Brauer08, Birnstiel10}. This prediction is supported by millimeter observations of disks \citep[][and references therein]{Testi14}. At this size, pebbles should experience fast radial drift due to the headwind they feel orbiting in the sub-Keplerian gas disk. This constant supply of drifting pebbles is a likely reservoir for planetesimal growth.

Assuming that radially drifting pebbles do indeed supply the inner disk with solids, the growth of the planetesimal disk boils down to how and where this material is halted from further migration.  \citet{Chatterjee14} proposed that migration of pebbles will be halted by a pressure bump in the inner disk associated with the inner edge of an MRI ``dead-zone". This would cause pebbles to collect in a dense ring at which point they would either become gravitationally unstable and collapse into protoplanets or they would undergo planet formation via core accretion. As this dead-zone boundary retreats, away from the star subsequent planet formation could occur, thus building STIPs from the inside out. An alternative solution is that pebbles simply grow fast enough and large enough in the inner disk that further radial drift is halted. \citet{Boley14} suggest that this rapid growth may be plausible due to a high concentration of solids in the inner disk and very efficient sticking of partially molten pebbles.

We propose an alternative means to stop pebbles from further migration.  We require that some seed planetesimals form in the inner disk. As we will show, the initial seed population can be very small. The rate of pebble accretion onto these existing planetesimals quickly surpasses the rate of new planetesimal formation in the inner disk such that the majority of mass in the final planetesimal disk is from the accretion of pebbles, not from the direct formation of new planetesimals. The appeal of this model is the generality of our assumptions. The formation of planetesimals almost certainly occurs as evidenced by the small bodies of the solar system which are presumably remnants of a planetesimal disk. It seems reasonable to extend this assumption to the inner disk as most models of planetesimal formation allow for the formation of planetesimals in this region \citep[e.g.][]{Chambers10b, Boley14, Carrera15}. However, uncertainties in the planetesimal formation process limit our ability to evaluate the validity of this assumption. The assumption of a supply of pebbles from the outer disk is consistent with both models and observation. Additionally, our model does not require the presence of a pressure bump or near perfect sticking between pebbles. Instead, high mass inner disks are a natural consequence of global accretion physics.


\section{Model}

The growth of a planetesimal can be described as:
\begin{equation}
\dot{M}_p = \pi b^2 \rho_{peb} v_{rel},
\label{3drate}
\end{equation}
where $b$ is the maximum distance a planetesimal can accrete a pebble from, $\rho_{peb}$ is the volume density of pebbles near the planetesimal and $v_{rel}$ is the relative velocity between the pebbles and the planetesimal. In the case where the accretion radius, $b$, is larger than the scale height of the pebble disk, this can be written as:
\begin{equation}
\dot{M}_p = 2b \Sigma_{peb} v_{rel},
\label{2drate}
\end{equation}
where $\Sigma_{peb}$ is the surface density of pebbles in the disk. In the following sections we describe how the density of pebbles in the disk and the accretion rate of these pebbles are determined.

\subsection{Protoplanetary disk profile}
The gas density throughout the disk is a key piece of information that is necessary for other parts of our model. We use a simple prescription for the gas surface density profile:
\begin{equation}
\Sigma_{gas} = \beta\left(\frac{r}{AU}\right)^{-p}.
\label{DiskEquation}
\end{equation}
We choose $p=1$, which is predicted for a steady-state viscous accretion disk \citep{Lynden74} and is also consistent with the range of values fitted to submillimeter observations of disks \citep{Andrews07}. Following \citet{Lambrechts14}, we set $\beta=\beta_0$exp$(-t/3$Myr) to account for the dissipation of the disk over time.
The gas volume density is related to the surface density by $\rho_{gas} \approx \Sigma_{gas}/2H$, where the scale height is given by:

\begin{equation}
\frac{H}{r} \approx 0.033\left(\frac{r}{AU}\right)^{1/4}.
\end{equation}

\subsection{Pebble surface density}
In our model, we assume that pebbles are continuously supplied from the outer disk and that there is no contribution to the pebble flux from within the simulation region ($<5$AU). The only sink for pebbles is by accretion onto planetesimals. The mass flux of pebbles through the disk at a given location is then:
\begin{equation}
F_{peb}(r) = F_{sup} - \sum_{r_p>r}\dot{M}_p =  2\pi r v_r\Sigma_{peb},
\end{equation}
where $F_{sup}$ is the supply rate of pebbles to the simulation region and the sum is over the accretion rates of all planetesimals outside of r. Inverting this equation gives us the local surface density of pebbles,
\begin{equation}
\Sigma_{peb} = \frac{F_{peb}}{2\pi r v_r}.
\label{surfacedensity}
\end{equation}

The timescale for pebbles to drift into the star from the outer disk tends to be much shorter than the timescale for pebble growth. Consequently, pebble growth can be considered the limiting factor in setting $F_{sup}$ and Equation \ref{surfacedensity} can be considered the instantaneous surface density.  \citet{Lambrechts14} have developed a model to describe the growth of pebbles and the evolution of the inward pebble mass flux over time, which we adopt for our work. For brevity, we do not repeat their derivation but instead, give their result. Assuming a gas disk profile as shown in Eq. \ref{DiskEquation}, they found the supply of pebbles over time to be

\begin{equation}
\begin{split}
F_{sup} = 9.5\times 10^{-5} & \left(\frac{\beta}{500g cm^{-2}}\right) \left(\frac{M_*}{M_{\odot}}\right)^{1/3} \left(\frac{Z_0}{0.01}\right)^{5/3} \\
& \left(\frac{t}{10^6 yr}\right)^{-1/3} M_{\oplus}yr^{-1}.
\label{pebblesupply}
\end{split}
\end{equation}
where $Z_0$ is the dust to gas mass ratio and $\beta$\, the gas surface density at 1AU, decays exponentially with time as in section 2.1.

At some point the supply of pebbles will cease. This will occur either when the disk has dissipated or when the pebble production front, as described in \citet{Lambrechts14}, moves beyond the outer radius of the protoplanetary disk. Assuming the typical extent of a disk is around 200 AU \citep{Andrews07}, the pebble supply would cease after 1-2 million years. We choose our simulation duration to be one million years as it is shorter than both the dissipation timescale and the pebble depletion timescale. This ensures that we do not supply more pebble mass to the inner disk than is available in the whole protoplanetary disk. It is possible that pebble accretion will occur longer than our simulation duration, but as we show later, this is similar to increasing the pebble production rate, which we do test.

The radial drift velocity of the pebbles is given by
\begin{equation}
v_r = -2 \frac{\rm{St}}{\rm{St}^2+1}\eta v_{kep}
\label{driftvelocity}
\end{equation}
\citep{Weidenschilling77b, Nakagawa86}, where $v_{kep}$ is the orbital Keplerian velocity. $\eta$ describes the pressure support on a body and can be estimated as
\begin{equation}
\eta = 0.0018\left(\frac{r}{AU}\right)^{1/2}.
\end{equation}
$\rm{St}=\tau_d \Omega$ is the Stokes number where $\Omega$ is the orbital frequency and the gas drag timescale, $\tau_d$, is
\begin{equation}
\begin{split}
\tau_d &= \frac{\rho_{m}s}{\rho_{gas}}  \;\;\;\;\;  s< \frac{9\lambda}{4} \;\;  (\rm{Epstein \;drag}) \\
&= \frac{4\rho_{m}s^2}{9\rho_{gas}c_s\lambda}  \;\;\;\;\;   \frac{9\lambda}{4} < s < \frac{27c_s\lambda}{2\eta v_{kep}}  \;\;(\rm{Stokes \;drag})\\
&=\frac{6\rho_{m}s}{\rho_{gas}\eta v_{kep}} \;\;\;\;\;  s > \frac{27c_s\lambda}{2\eta v_{kep}} \;\;(\rm{Quadratic \;drag}).
\end{split}
\end{equation}
$\rho_m$ is the material density of the pebble, $s$ is the size of the pebble, $c_s$ is the local sound speed of the gas and $\lambda=2\times 10^{-9}/\rho_{gas}$ \citep[cgs units;][]{Nakagawa86} is the mean free path of the gas.

Substitution of Eqs.\  \ref{pebblesupply} and \ref{driftvelocity} into Eq.\ \ref{surfacedensity} yields the surface density of pebbles throughout the disk over time. The pebble volume density can be approximated by dividing the surface density by 2 times the scale height of the pebbles:
\begin{equation}
H_{peb} = H_{gas}\left(\frac{\alpha}{\alpha+\rm{St}}\right)^{1/2}
\label{pebblescaleheight}
\end{equation}
\citep{Youdin07}, where $\alpha$ is the turbulent diffusion parameter for a viscous accretion disk \citep{Shakura73}.

\subsection{Pebble Accretion Rate}

To obtain the accretion rates of planetesimals, we need to know their collisional cross-section. In the simplest case, this is the geometrical cross-section. In reality, other factors act to enhance the cross-section above the geometrical limit such as gravitational focusing and gas drag. Gas drag, in particular, is critically important when considering the accretion of small pebbles because its affect on the trajectories of pebbles is of comparable magnitude as the affect of gravity. \citet{Ormel10} calculated the trajectories of small bodies near planetesimals and found that the combined influence of gravity and gas drag on the small bodies can increase a planetesimal's accretional cross-section significantly - up to a large fraction of the planetesimal's hill radius. Subsequent studies have employed this result to help explain the short timescales that are necessary to form gas giant cores before the dispersal of the gas disk \citep{Lambrechts12, Chambers14, Lambrechts14, Kretke14}.

In our work, we apply the results of \citet{Ormel10} to the accretion of pebbles in the inner disk. We direct the reader to their paper for details of these calculations. In particular, their Table 2 provides a summary of the steps we took to determine the collisional cross-section of planetesimals.

The remaining term that is needed to determine the accretion rate of pebbles is the relative velocity between the planetesimal and the pebbles. We take this to be the larger of either the radial drift velocity (Eq.\ \ref{driftvelocity}) or the velocity imparted upon the pebbles due to the turbulent motions of the gas,
\begin{equation}
v_{turb}^2 = \alpha c_s^2 \times \rm{min}\left(2\rm{St}, \frac{1}{1+\rm{St}}\right)
\end{equation}
\citep{Ormel07, Chambers14}.

The accretion rate is calculated first for the outermost planetesimal then proceeding inward to the innermost. The surface density of pebbles at a planetesimal location is calculated according to Equation \ref{surfacedensity}. The accretion rate is then calculated by inserting this value, the relative velocity between the planetesimal and pebbles and the calculated accretion radius into Equations\ \ref{3drate} or \ref{2drate}. The inward pebble flux is then decreased by the accretion rate of that planetesimal in order to conserve mass and the process is repeated for the next planetesimal.

\subsection{Initial Conditions}

 We begin the simulations with a low mass disk of planetesimals. Conceptually, the start of our simulations corresponds to the point at which the rate of pebble accretion exceeds the rate of new planetesimal formation. Because, the process of planetesimal formation is not well understood, the distribution of the first planetesimals is poorly constrained. For this reason, we test a range of initial planetesimal distributions with the following function form:
\begin{equation}
\Sigma_p = \Sigma_{0}\left(\frac{a}{AU}\right)^{\alpha}.
\end{equation}
 The locations of the planetesimals are determined by assuming a surface density profile for the planetesimal disk. Planetesimals, with some size range/distribution, are then placed in the disk so that they follow the assumed surface density profile.

 The expected sizes of newly born planetesimals depends on how they form. If they form via pairwise accretion then a continuum of sizes from small to large would be expected. If instead they form from the gravitational collapse of pebbles in either turbulent concentrations \citep{Cuzzi08, Chambers10b} or streaming instabilities \citep{Johansen07} then we would expect the initial size distribution of planetesimals to be larger. The size distribution of bodies in the asteroid belt also supports the idea that planetesimals (at least in that region of the disk) were born big \citep{Bottke05}. In our simulations, we also assume that planetesimals form large. For most of our simulations, the initial planetesimal sizes follow a distribution similar to the inferred initial size distribution of the asteroid belt \citep{Bottke05}:
 \begin{equation}
 N(>D) \propto D^{-4.5} \;\;\;\;\; 100\;\rm{km} < D < 500\;\rm{km}.
 \end{equation}


\section{Simulations}

\begin{table}
 \caption{Simulation paramters}
\begin{tabular}{l l p{4.5cm}}
\hline
\hline
Parameter 	& Default value		& Description \\
\hline
$\alpha_{v}$	& 0.001				& Viscosity parameter \\
s			& 1 cm				& Pebble size \\
$\rho_{peb}$	& 2 $\rm{g\, cm}^{-3}$	& Pebble material density \\
R$_{p}$		& 100-500 km			& Planetesimal size \\
$\rho_{p}$		& 3 $\rm{g\, cm}^{-3}$	& Planetesimal material density \\
$\beta_{0}$	& 500 $\rm{g\, cm}^{-2}$ 	& Initial gas surface density at 1 AU \\
$\Sigma_{0}$	& 1 $\rm{g\, cm}^{-2}$	& Initial planetesimal surface density at 1 AU\\
$\alpha$		& -1.5				& Initial planetesimal surface density power law index \\
\hline
\end{tabular}

\label{paramlist}

\end{table}

\begin{figure}     
\epsscale{1}      
\plotone{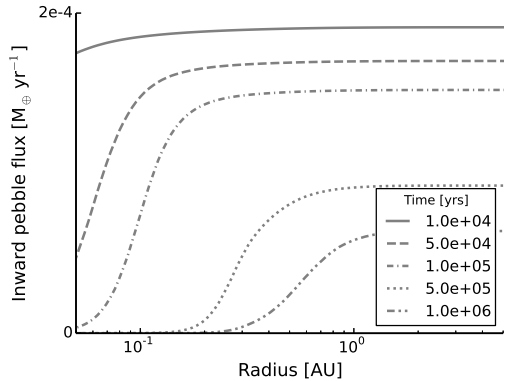}
\caption{Inward flux of pebbles throughout the disk at different times during the simulation. The reduction in pebble flux closer to the star is due to pebble accretion by planetesimals.}
\label{inward_flux}
\end{figure}

\begin{figure}     
\epsscale{1}      
\plotone{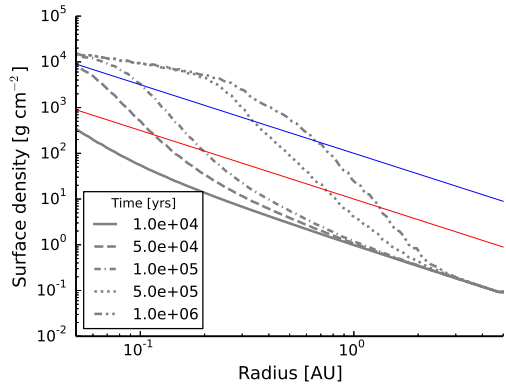}
\caption{Surface density of planetesimals throughout the disk at different times during the simulation. Red and blue lines correspond to the MMSN and 10 times the MMSN respectively.}
\label{baseline_sd}
\end{figure}

\subsection{Baseline Simulation}
We begin by examining the general behavior of the accretion simulations for a baseline case. Our model parameters and their values for this baseline case are listed in Table \ref{paramlist}. In the following sections we explore how the choice of parameter values affects the the growth of the planetesimal disk.

A striking feature of this simulation is the high efficiency with which pebbles are accreted onto planetesimals. Figure \ref{inward_flux} shows the inward flux of pebbles through the disk as a function of semi-major axis for a number of different times in the simulation. Except at the very beginning of the simulation, nearly all pebbles are accreted before they can make it through the disk. We will refer to the earliest time at which all pebbles are accreted as the time of total accretion.

The pebble accretion rate is a strong function of location in the disk. The reason for this is twofold. The surface density of planetesimals increases closer to the star. Therefore, any particular pebble drifting in through the disk has a higher probability of being accreted by a planetesimal the closer it gets to the star. The second reason is that as pebbles drift inwards they are concentrated from a relatively large volume into one much smaller thus increasing the space density of pebbles dramatically at small radii. Planetesimals accreting from this concentrated flux of pebbles will grow rapidly. Because the accretion rate is such a strong function of distance from the star, the radial transition in the disk from minimal accretion to significant accretion is sharp. This edge of accretion moves out in the disk over time as planetesimals farther out in the disk grow large enough to begin accreting significantly from the pebble flux. Interior to this edge the accretion rate begins to drop off again. This decrease is due to the fact that most of the pebbles have already been accreted by planetesimals farther out and so the supply is much reduced. This forms an inner edge to the accretion. Accretion within this annulus accounts for the majority of pebble accretion throughout the disk at any given time. As the accretion annulus progresses outward over time, the planetesimal disk is built up from the inside out, leading to the evolution depicted in Figure \ref{baseline_sd}.

\subsection{Initial Planetesimal Size}

	The initial sizes of planetesimals are not known. Evidence from the asteroid belt suggests that, in that region of the disk, planetesimals were born large. For most of our simulations we begin with a distribution of planetesimal sizes that are consistent with that inferred from the asteroid belt \citep{Bottke05}. However, we do not know if this size distribution is accurate for the rest of the disk. Here we consider the effects of changing the initial sizes of the planetesimals.

\begin{figure}     
\epsscale{1}      
\plotone{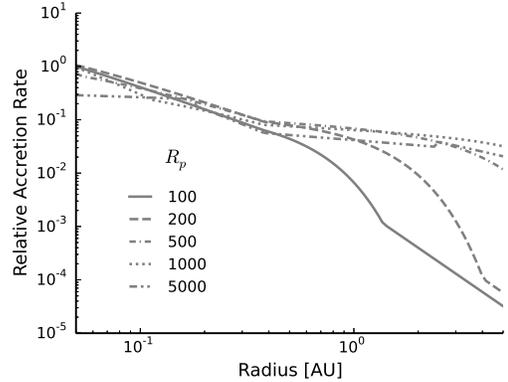}
\caption{Relative accretion rate of pebbles for disks composed of different sizes ($R_p$ in units of km) of planetesimals. Each disk has the same total mass and radial surface density profile.}
\label{accretion_vs_planetesimalsize}
\end{figure}

	The importance of planetesimal size is best demonstrated by examining the results of \citet{Ormel10}. In their work, they numerically integrated the trajectories of small particles near planetesimals and included the effects of gas drag. The combined effects of gravity and gas drag lead to a somewhat complicated dependence of the accretional cross-section on planetesimal size, pebble size and disk properties. In Figure \ref{accretion_vs_planetesimalsize}, we calculate the pebble accretion rate, using the collisional cross-sections from \citep{Ormel10}, throughout the disk for disks composed of planetesimals of different sizes. The total pebble accretion rate in the inner disk depends weakly on the size of the planetesimals except at very large sizes at which point accretion becomes less efficient. Beyond $\sim$ 1 AU, a disk composed of larger planetesimals accretes significantly faster than one composed of smaller planetesimals.

\begin{figure}     
\epsscale{1}      
\plotone{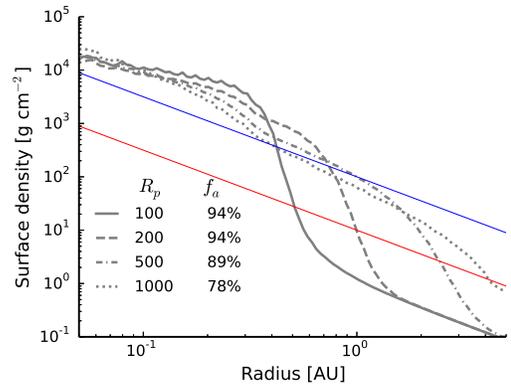}
\caption{Final surface density of the planetesimal disk in simulations starting with different initial planetesimal sizes ($R_p$ in units of km). Red and blue lines correspond to the MMSN and 10 times the MMSN respectively. $f_a$ gives the fraction of pebbles that were accreted during the simulation.}
\label{sd_vs_planetesimalsize}
\end{figure}

With this insight, we can interpret the results of the accretion simulations with different initial planetesimal sizes. Figure \ref{sd_vs_planetesimalsize} shows the final surface density of the planetesimal disk for simulations with four different initial planetesimal sizes. As might be expected, disks with larger planetesimals accrete more pebbles farther out in the disk. This leads to a reduced pebble flux closer to the star, and, consequently, a less massive inner disk. Disks composed of larger planetesimals are also less efficient at accreting planetesimals over the course of the full simulation. This is because planetesimals larger than $\sim$ 2000 km are less efficient (per unit mass) at accreting pebbles. The initially larger planetesimals will cross this size threshold sooner and therefore accrete less efficiently over their lifetime.

\subsection{Initial Planetesimal Distribution}
The initial placement and total mass of planetesimals in the disk are perhaps the most ill-constrained aspects of these simulations. In reality, the distribution of the first planetesimals must depend on the details of planetesimal formation and how they change throughout the disk. Until we have a better understanding of planetesimal formation we must base our initial conditions on other information.

Observations of gas in protoplanetary disks provide information on the distribution of mass prior to and during planetesimal formation.  \citet{Andrews07} fit power laws to the gas surface density as a function of disk radius and found the power law index to range between 0 and -2. On the other end, observations of planetary systems provide us with the final distribution of mass once the planet formation process has completed. On the whole, exoplanetary systems have a surface density power law index of -1.5 \citep{Chiang13}, consistent with the MMSN, but individual systems have a wide range of power law indices \citep{Raymond14}. These constraints provide us with a starting point, but it is important to keep in mind that they are not strict constraints as the solid mass distribution in the intermediate stages of planet formation does not necessarily reflect the mass distribution in the initial and final stages.

\begin{figure}     
\epsscale{1}      
\plotone{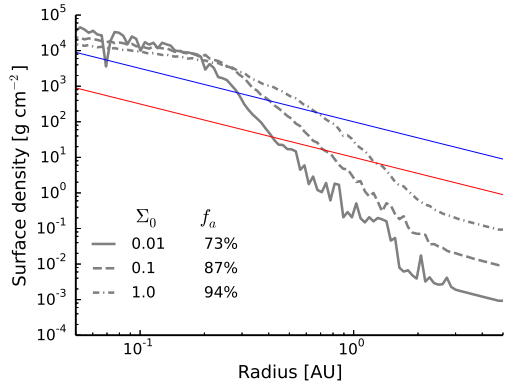}
\caption{Final surface density of the planetesimal disk in simulations starting with different initial total masses in planetesimals. $\Sigma_0$ is the initial surface density of planetesimals at 1 AU in units of $g\,cm^{-2}$. Red and blue lines correspond to the MMSN and 10 times the MMSN respectively. $f_a$ gives the fraction of pebbles that were accreted during the simulation.}
\label{sd_vs_planetesimalmass}
\end{figure}

\begin{figure}     
\epsscale{1}      
\plotone{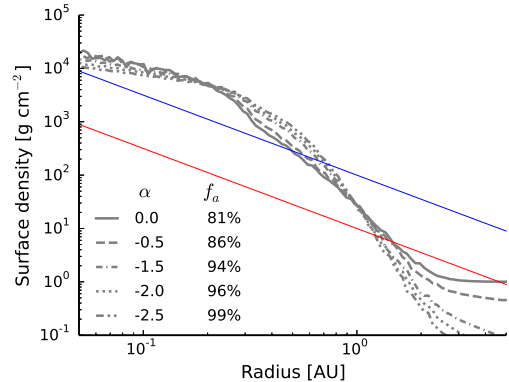}
\caption{Final surface density of the planetesimal disk in simulations starting with initial planetesimal surface density profiles with different power law indices ($\alpha$). Red and blue lines correspond to the MMSN and 10 times the MMSN respectively. $f_a$ gives the fraction of pebbles that were accreted during the simulation. }
\label{sd_vs_slope}
\end{figure}

In most of our simulations, we began with the initial planetesimal distribution following a tenth of the MMSN. Here we remove this restriction and allow both the total mass in planetesimals and the slope of the mass distribution to vary. Figure \ref{sd_vs_planetesimalmass} shows the final planetesimal surface density for three different simulations beginning with a planetesimal surface density distribution 0.001, 0.01 and 0.1 times the MMSN. The effects of changing the initial amount of mass in the system are relatively limited. Simulations with initially lower mass planetesimal disks are less efficient at accreting pebbles over the course of the simulation. Nevertheless, even the very low mass disk accretes the majority of pebbles. This difference can be attributed to the initially lower mass disk taking longer to reach the time of total accretion (time where all pebbles are accreted) and the relatively small magnitude of this difference can be attributed to the time of total accretion being much less than the length of the simulation.
Similarly, changing the slope of the initial planetesimal surface density profile has little impact on the final surface density distribution of planetesimals (see Figure \ref{sd_vs_slope}.

 An interesting feature is that the simulation with the initially lower mass disk ends up with a slightly higher surface density in the innermost regions than the other simulations. This results because the accretion annulus moves out in the disk more slowly resulting in a longer period of accretion in the inner disk. On the other hand, this results in a shorter period of accretion for the outer portions of the disk and a consequently lower final surface density in the outer disk.

\subsection{Pebble Size}
\begin{figure}     
\epsscale{1}      
\plotone{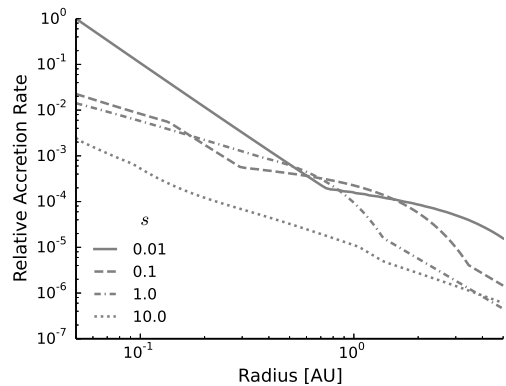}
\caption{Relative accretion rate of pebbles of different sizes ($s$ in units of cm) throughout the disk by 200 km planetesimals. In this figure, unlike the simulations, all parts of the disk receive the same radial pebble flux. }
\label{accretion_vs_pebblesize}
\end{figure}

The size of the pebbles being accreted has a significant effect on the accretion rate of a planetesimal. The typical size of pebbles in the inner regions of disks is unknown, but we can test a range of values that are motivated from dust coagulation simulations. Such simulations find the dominant particle size to range from 0.01-10 cm \citep[e.g.][]{Birnstiel10, Brauer08}.  Figure \ref{accretion_vs_pebblesize} demonstrates the difference in accretion rate throughout the disk if we vary the pebble size over this range. It is clear that the accretion rate's dependance on pebble size is complicated and even more so if we consider that the size of the accreting planetesimal also changes the pebble size dependance.

\begin{figure}     
\epsscale{1}      
\plotone{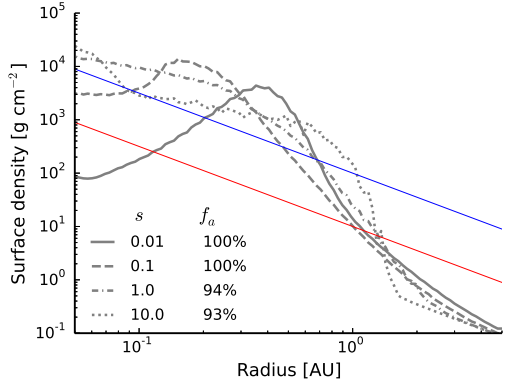}
\caption{Final surface density of the planetesimal disk in simulations with different pebble sizes ($s$ in units of cm). Red and blue lines correspond to the MMSN and 10 times the MMSN respectively. $f_a$ gives the fraction of pebbles that were accreted during the simulation.}
\label{sd_vs_pebblesize}
\end{figure}

When we vary the pebble size in our simulations, we see a corresponding change in the final planetesimal surface density (Figure \ref{sd_vs_pebblesize}). Somewhat counter-intuitively, the smaller pebbles, which tend to be more efficiently accreted, result in less massive inner disks. Once again it is easiest to think in terms of the movement of the accretion annulus. In simulations with small pebbles, this annulus moves out in the disk rather rapidly so that the innermost regions do not have time to accrete a lot of mass. This is then compensated by more accretion farther out. In all cases the efficiency of pebble accretion for the disk as a whole is very efficient, it is just the distribution of this mass that changes as a function of pebble size.

\subsection{Turbulence}
\begin{figure}     
\epsscale{1}      
\plotone{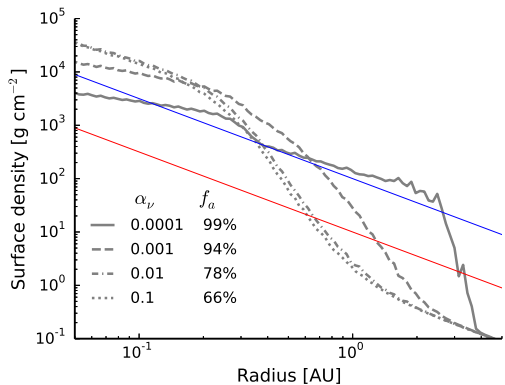}
\caption{Final surface density of the planetesimal disk in simulations with different turbulent diffusion parameters ($\alpha_{\nu}$). Red and blue lines correspond to the MMSN and 10 times the MMSN respectively. $f_a$ gives the fraction of pebbles that were accreted during the simulation..}
\label{sd_vs_turbulence}
\end{figure}

Turbulence in the disk acts to increase the scale height of pebbles in the disk (Eq.\ \ref{pebblescaleheight}). An increase in scale height corresponds to a decrease in the volume density of pebbles and consequently a reduction in the efficiency of accretion. A reduction in accretion efficiency slows the outward movement of the accretion annulus resulting in more mass being accreted closer to the star (see Figure \ref{sd_vs_turbulence}). Increasing the strength of turbulence does reduce the efficiency of pebble accretion, but not enough to prevent the formation of massive planetesimal disks.   Instead it has more of an impact on where in the disk this mass is accreted.

\subsection{Gas Mass}

One of the more interesting parameters to examine is the total mass of the disk. All the previous parameters we examine, we varied to account for their uncertainty. In reality these parameters may be very similar in different disks. Disk mass, on the other hand, is observed to vary from disk to disk and consequently has some predictive power with regards to the distribution of mass in planetary systems.

Varying the mass of the disk in our model is effectively varying the supply rate of pebbles - a higher mass disk leads to a larger supply of pebbles. The dust to gas ratio also determines the pebble supply rate so these two parameters are mostly degenerate. We will restrict our discussion to the total disk mass, but the same effects can be achieved by varying the dust to gas ratio in a corresponding manner.

\begin{figure}     
\epsscale{1}      
\plotone{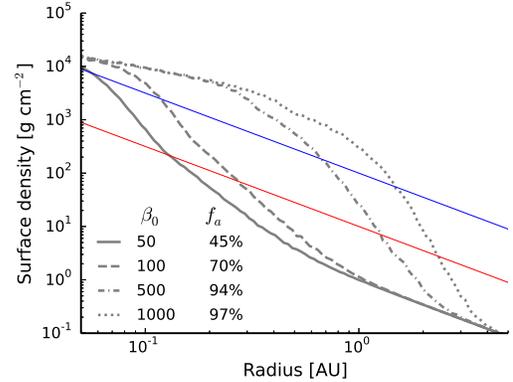}
\caption{Final surface density of the planetesimal disk in simulations with different initial gas disk masses. The parameter, $\beta_0$, is the initial gas surface density at 1 AU in units of $g\,cm^{-2}$. Red and blue lines correspond to the MMSN and 10 times the MMSN respectively. $f_a$ gives the fraction of pebbles that were accreted during the simulation.}
\label{sd_vs_diskmass}
\end{figure}

The effect of changing the disk mass is relatively straight-forward. A lower mass disk essentially slows down the whole accretion sequence. The accretion annulus moves out in the disk much more slowly. This results a lower mass planetesimal disk and a larger fraction of the accreted mass lying in the innermost regions of the disk (Figure \ref{sd_vs_diskmass}). Note the similarity between the evolutionary sequence of the growing planetesimal disk shown in Figure \ref{baseline_sd} and the different mass disks of Figure \ref{sd_vs_diskmass}. In this sense, changing the disk mass has a similar effect to suddenly cutting off the supply of pebbles to the inner disk.

\subsection{Concurrent planetesimal formation and accretion}

A common feature in the results of many of the previous simulations is a lack of mass in the planetesimal disk at more distant radii. This is a consequence of the longer planetesimal growth timescales and at these distances. In all of these simulations, we began with some amount of mass in planetesimals and assumed that the pebble accretion rate was much larger than the rate of new planetesimal formation (and thus did not include any further planetesimal formation). However, this is clearly not the case for the outer disk where planetesimal formation must continue to be significant in order to produce enough mass to match, for example, the MMSN.

To address this problem, we ran a simulation that continues to form planetesimals throughout the course of the simulation. Because the rate of planetesimal formation in disks is unknown, we take the simplest case and assume that it is constant in time for the extent of the simulation. We start with an empty disk and at each timestep add planetesimals to the disk such that by the end of the simulation, were there no pebble accretion, the planetesimal disk would have a surface density profile corresponding to the MMSN.

\begin{figure}     
\epsscale{1}      
\plotone{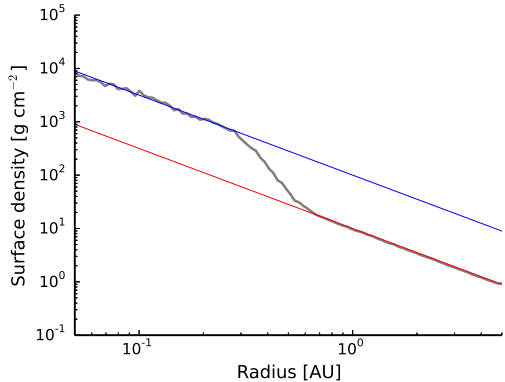}
\caption{Final surface density of the planetesimal disk in a simulation that included pebble accretion and planetesimal formation throughout the entire simulation. The model parameters are the same as those in Table 1 except that this simulation begins with no planetesimal disk. Red and blue lines correspond to the MMSN and 10 times the MMSN respectively. }
\label{continuous_formation}
\end{figure}

The results of this simulation can be seen in Figure \ref{continuous_formation}. Not surprisingly, the surface density profile at larger radii, where pebble accretion was negligible, matches the MMSN. In the inner disk, where pebble accretion was not negligible, the surface density is ten times higher than the MMSN, consistent with our previous simulations. This simulation is likely more realistic than the previous simulations because planetesimal formation and pebble accretion probably occur concurrently. However, we do not explore this scenario further because of the large uncertainties in the rate of planetesimal formation. The combination of planetesimal formation and pebble accretion throughout the lifetime of the disk is a realistic way to form a massive inner disk without causing an underdeveloped outer disk. This allows for the formation of STIPs and the formation of systems like the solar system from the same model, albeit with the requirement that some other mechanism eventually remove the inner disk material \citep[e.g.][]{Batygin15, Pu15, Volk15}

\section{Discussion}
The most prominent result from our simulations is that the accretion of pebbles in the inner disk is an efficient process. Furthermore, this efficiency depends weakly on the model parameters. In all but one simulation (the very low mass protoplanetary disk simulation), the efficiency of pebble accretion was greater than 50\% and in most cases much higher than that.  Our results suggest that high mass and compact planetesimal disks are the expected outcome of planetesimal formation and growth in protoplanetary disks.

\subsection{Variation between planetary systems}
The processes described by our model are expected to operate in all protoplanetary disks. The observed distribution of planets shows a diversity of planetary system architectures. Our model naturally allows for this variability of observed systems.

The main source of differences between systems in our model is the extent of accretion. This is controlled by the duration of accretion and the rate of pebble delivery. Planetesimal disks that have more time to grow or receive a larger pebble flux are more massive and they are less centrally condensed than disks that have less time to grow or receive lower pebbles fluxes (see Figures \ref{baseline_sd} \& \ref{sd_vs_diskmass}).

Planetesimals will continue to accrete until the supply of pebbles is shut off. This may last as long as it takes for the disk to dissipate or in the event of giant planet formation, it may have a much shorter duration. A giant planet in the outer disk will accrete the vast majority of pebbles drifting by thus significantly reducing the pebble flux to the inner disk. If giant planet formation occurs early on (within 1-2 million years), when pebble accretion is growing the inner disk, then we would expect that, on average, systems containing giant planets would have a lower total mass in terrestrial planets and that the terrestrial planet mass would be concentrated in shorter period orbits. Individual systems containing giant planets would likely show a continuum of architectures depending on the exact timing of giant planet formation.

The total flux in pebbles to the inner disk depends on two main factors: the total mass of the disk and the solid fraction of the disk. Observations of disks \citep[e.g.][]{Andrews07} reveal a broad range of protoplanetary disk masses spanning several orders of magnitude. Similarly, the span of metallicities found in planet hosting stars \citep[e.g.][]{Valenti05} implies a broad range in dust to gas ratios in the disks from which they formed. These large ranges in disk mass and gas to dust ratios would lead to significant variations in the supply rate of pebbles in different disks and consequently to variations in the structure of the planetesimal disk.

Protoplanetary disk mass has been shown to correlate with stellar type \citep{Andrews13}. We might then expect planetesimal disks around larger stars to be more massive and less compact. However, the formation of giant planets complicates matters. Giant planet formation has also been shown to correlate with stellar mass \citep{Johnson07}. Therefore the expected trend with stellar mass may be washed out or removed due to the increased frequency of giant planets. Similarly, giant planet occurrence rates are known to correlate with host star metallicity \citep[e.g.][]{Gonzalez97, Santos01, Fischer05, Buchhave14, Wang15} potentially negating possible trends with metallicity.

Our simulations also showed that changing other parameters, such as the initial planetesimal sizes or the turbulent diffusion parameter, changes the resulting density distribution of the planetesimal disk. The range of parameter values that we tested were meant to reflect their uncertainty and not their intrinsic variation between protoplanetary disks. However, some variation may still occur between disks, especially in disks around different type stars leading to additional variability in planetesimal disk structure.

In summary, the variation in protoplanetary disk masses and gas to dust ratios and the variation in the duration of pebble accretion give rise to variations in planetesimal disk structure. These factors affect the surface density structure of the planetesimal disk in similar ways and may, in some cases, counteract each other. We predict a trend of increasing planetesimal disk mass and decreasing compactness with stellar mass and metallicity and the lack of a giant planet in the system. However, these trends may be diluted due to the increased frequency of giant planets around more massive and metal-rich stars. On the whole, the range of planetesimal disk structures predicted by our model can account for the diversity of observed planetary systems.

\subsection{Final assembly of planets}
The logical next step is to see what kind of planetary systems will assemble from the planetesimal disks formed in these simulations. This final step is, as of yet, an unsolved problem, but we will comment on the implications of our results. This work removes a large hurdle for the scenario in which STIPs form in situ. Our results show that high mass and compact planetesimal disks are not just possible but are the expected outcome of planetesimal formation and growth. Thus, if we ignore the effects of migration, STIPs will naturally form from in situ accretion in these disks.

If, however, we do not ignore migration, do STIPs still form? This is a more difficult question to answer because simulations including migration have not relied on a massive inner disk and therefore did not include one. Additional work will need to be done to see if it is possible to grow planets farther out and migrate them through a massive planetesimal disk to form STIPs. Another possibility is that planets accrete most of their mass as they move through the inner disk.

Regardless of how the final assembly of planets occurs, an interesting question to be addressed in future work is: does the range of planetesimal disk surface density profiles expected from these simulations produce planetary systems with the range of observed architectures? Examining the more detailed statistics of systems produced from simulations may be our best bet in determining how planets are ultimately pieced together.


\section{Summary and Conclusions}

We have put forward a model for the growth of a planetesimal disk that is, in principle, straight-forward - planetesimal disks are built from the combination of direct planetesimal formation and the accretion of pebbles onto existing planetesimals. This model rests on two main assumptions: planetesimal formation occurs throughout the disk and pebbles rapidly grow in the outer disk and then drift through the inner disk.

In practice, our model is more complicated due to the complexity of pebble accretion and protoplanetary disk structure. This leads to a large number of model parameters, with large uncertainties, that affect the growth pattern of the planetesimal disk. However, one aspect that is common to our simulations regardless of these parameter values is that pebble accretion is efficient in the inner disk. This high efficiency allows for the formation of high mass inner disks without the need for a massive protoplanetary disk or enhanced planetesimal formation rates.

The final structure of planetesimal disks in our simulations is controlled by the rate of pebble delivery to the inner disk and the duration of this delivery. More massive protoplanetary disk and those with high dust to gas ratios will supply more pebbles to the inner disk and consquently build more massive planetesimal disks. The duration of pebble delivery is controlled by the lifetime of the protoplanetary disk and potentially the timing of giant planet formation. Giant planets will act as road blocks to any inward migrating pebbles thus starving the inner disk. With these comments in mind, we expect that more massive and/or metal-rich stars (with more massive disks and higher dust to gas ratios) and stars that do not host giant planets will contain more mass in terrestrial planets.

Exactly how planetesimal disks evolve into planetary systems is still a matter of debate. However, our results strongly suggest that the initial conditions for late stage planet formation should include a massive planetesimal disk. This does support the idea that STIPs could form in situ without the need for the migration of material during or after planet assembly. However, this does not preclude the possibility that close in planets migrated significantly to their final orbits.

\acknowledgments
The authors thank the referee for useful comments used to improve the paper. This material is based upon work supported by NASA under award No. NNX15AF02G.


\begin{thebibliography}{42}
\expandafter\ifx\csname natexlab\endcsname\relax\def\natexlab#1{#1}\fi

\bibitem[{Amelin {et~al.}(2002)Amelin, Krot, Hutcheon, \& Ulyanov}]{Amelin02}
Amelin, Y., Krot, A.~N., Hutcheon, I.~D., \& Ulyanov, A.~A. 2002, Science, 297,
  1678

\bibitem[{Andrews {et~al.}(2013)Andrews, Rosenfeld, Kraus, \&
  Wilner}]{Andrews13}
Andrews, S.~M., Rosenfeld, K.~A., Kraus, A.~L., \& Wilner, D.~J. 2013, ApJ,
  771, 129

\bibitem[{Andrews \& Williams(2007)}]{Andrews07}
Andrews, S.~M., \& Williams, J.~P. 2007, ApJ, 659, 705

\bibitem[{Batygin \& Laughlin(2015)}]{Batygin15}
Batygin, K., \& Laughlin, G. 2015, in PNAS, 4214--4217

\bibitem[{Batygin \& Morbidelli(2013)}]{Batygin13}
Batygin, K., \& Morbidelli, A. 2013, AJ, 145, 1

\bibitem[{Birnstiel {et~al.}(2010)Birnstiel, Dullemond, \&
  Brauer}]{Birnstiel10}
Birnstiel, T., Dullemond, C.~P., \& Brauer, F. 2010, A{\&}A, 513, A79

\bibitem[{Boley {et~al.}(2014)Boley, Morris, \& Ford}]{Boley14}
Boley, A.~C., Morris, M.~A., \& Ford, E.~B. 2014, ApJL, 792, L27

\bibitem[{Bottke {et~al.}(2005)Bottke, Durda, Nesvorny, Jedicke, Morbidelli,
  Vokrouhlick{\'y}, \& Levison}]{Bottke05}
Bottke, W.~F., Durda, D.~D., Nesvorny, D., {et~al.} 2005, Icarus, 175, 111

\bibitem[{Brauer {et~al.}(2008)Brauer, Dullemond, \& Henning}]{Brauer08}
Brauer, F., Dullemond, C.~P., \& Henning, T. 2008, A{\&}A, 480, 859

\bibitem[{Buchhave {et~al.}(2014)Buchhave, Bizzarro, Latham, Sasselov, Cochran,
  Endl, Isaacson, Juncher, \& Marcy}]{Buchhave14}
Buchhave, L.~A., Bizzarro, M., Latham, D.~W., {et~al.} 2014, Nature, 509, 593

\bibitem[{Carrera {et~al.}(2015)Carrera, Johansen, \& Davies}]{Carrera15}
Carrera, D., Johansen, A., \& Davies, M.~B. 2015, A{\&}A, 579, A43

\bibitem[{Chambers(2010)}]{Chambers10b}
Chambers, J.~E. 2010, Icarus, 208, 505

\bibitem[{Chambers(2014)}]{Chambers14}
---. 2014, Icarus, 233, 83

\bibitem[{Chatterjee \& Tan(2014)}]{Chatterjee14}
Chatterjee, S., \& Tan, J.~C. 2014, ApJ, 780, 53

\bibitem[{Chiang \& Laughlin(2013)}]{Chiang13}
Chiang, E., \& Laughlin, G. 2013, MNRAS, 431, 3444

\bibitem[{Cuzzi {et~al.}(2008)Cuzzi, Hogan, \& Shariff}]{Cuzzi08}
Cuzzi, J.~N., Hogan, R.~C., \& Shariff, K. 2008, ApJ, 687, 1432

\bibitem[{Dressing {et~al.}(2015)Dressing, Charbonneau, Dumusque, Gettel, Pepe,
  Collier-Cameron, Latham, Molinari, Udry, Affer, Bonomo, Buchhave, Cosentino,
  Figueira, Fiorenzano, Harutyunyan, Haywood, Johnson, Lopez-Morales, Lovis,
  Malavolta, Mayor, Micela, Motalebi, Nascimbeni, Phillips, Piotto, Pollacco,
  Queloz, Rice, Sasselov, S{\'e}gransan, Sozzetti, Szentgyorgyi, \&
  Watson}]{Dressing15}
Dressing, C.~D., Charbonneau, D., Dumusque, X., {et~al.} 2015, ApJ, 800, 135

\bibitem[{Fischer \& Valenti(2005)}]{Fischer05}
Fischer, D.~A., \& Valenti, J. 2005, ApJ, 622, 1102

\bibitem[{Gonzalez(1997)}]{Gonzalez97}
Gonzalez, G. 1997, MNRAS, 285, 403

\bibitem[{Hands {et~al.}(2014)Hands, Alexander, \& Dehnen}]{Hands14}
Hands, T.~O., Alexander, R.~D., \& Dehnen, W. 2014, MNRAS, 445, 749

\bibitem[{Hansen \& Murray(2012)}]{Hansen12}
Hansen, B. M.~S., \& Murray, N. 2012, ApJ, 751, 158

\bibitem[{Johansen {et~al.}(2007)Johansen, Oishi, Low, Klahr, Henning, \&
  Youdin}]{Johansen07}
Johansen, A., Oishi, J.~S., Low, M.-M.~M., {et~al.} 2007, Nature, 448, 1022

\bibitem[{Johnson {et~al.}(2007)Johnson, Butler, Marcy, Fischer, Vogt, Wright,
  \& Peek}]{Johnson07}
Johnson, J.~A., Butler, R.~P., Marcy, G.~W., {et~al.} 2007, ApJ, 670, 833

\bibitem[{Kley \& Nelson(2012)}]{Kley12}
Kley, W., \& Nelson, R.~P. 2012, ARA{\&}A, 50, 211

\bibitem[{Kretke \& Levison(2014)}]{Kretke14}
Kretke, K.~A., \& Levison, H.~F. 2014, AJ, 148, 109

\bibitem[{Lambrechts \& Johansen(2012)}]{Lambrechts12}
Lambrechts, M., \& Johansen, A. 2012, A{\&}A, 544, A32

\bibitem[{Lambrechts \& Johansen(2014)}]{Lambrechts14}
---. 2014, A{\&}A, 572, A107

\bibitem[{Lithwick \& Wu(2012)}]{Lithwick12}
Lithwick, Y., \& Wu, Y. 2012, ApJL, 756, L11

\bibitem[{Lynden-Bell \& Pringle(1974)}]{Lynden74}
Lynden-Bell, D., \& Pringle, J.~E. 1974, MNRAS, 168, 603

\bibitem[{Nakagawa {et~al.}(1986)Nakagawa, Sekiya, \& Hayashi}]{Nakagawa86}
Nakagawa, Y., Sekiya, M., \& Hayashi, C. 1986, Icarus, 67, 375

\bibitem[{Ormel \& Cuzzi(2007)}]{Ormel07}
Ormel, C.~W., \& Cuzzi, J.~N. 2007, A{\&}A, 466, 413

\bibitem[{Ormel \& Klahr(2010)}]{Ormel10}
Ormel, C.~W., \& Klahr, H.~H. 2010, A{\&}A, 520, A43

\bibitem[{Pu \& Wu(2015)}]{Pu15}
Pu, B., \& Wu, Y. 2015, arXiv:150205449

\bibitem[{Raymond \& Cossou(2014)}]{Raymond14}
Raymond, S.~N., \& Cossou, C. 2014, MNRASL, 440, L11

\bibitem[{Santos {et~al.}(2001)Santos, Israelian, \& Mayor}]{Santos01}
Santos, N.~C., Israelian, G., \& Mayor, M. 2001, A{\&}A, 373, 1019

\bibitem[{Shakura \& Sunyaev(1973)}]{Shakura73}
Shakura, N.~I., \& Sunyaev, R.~A. 1973, A{\&}A, 24, 337

\bibitem[{Terquem \& Papaloizou(2007)}]{Terquem07}
Terquem, C., \& Papaloizou, J. C.~B. 2007, ApJ, 654, 1110

\bibitem[{Testi {et~al.}(2014)Testi, Birnstiel, Ricci, Andrews, Blum,
  Carpenter, Dominik, Isella, Natta, Williams, \& Wilner}]{Testi14}
Testi, L., Birnstiel, T., Ricci, L., {et~al.} 2014, in Protostars and Planets
  VI (University of Arizona Press), 339--361

\bibitem[{Valenti \& Fischer(2005)}]{Valenti05}
Valenti, J.~A., \& Fischer, D.~A. 2005, ApJS, 159, 141

\bibitem[{Volk \& Gladman(2015)}]{Volk15}
Volk, K., \& Gladman, B. 2015, arXiv:150206558

\bibitem[{Wang \& Fischer(2015)}]{Wang15}
Wang, J., \& Fischer, D.~A. 2015, AJ, 149, 14

\bibitem[{Weidenschilling(1977)}]{Weidenschilling77b}
Weidenschilling, S.~J. 1977, MNRAS, 180, 57

\bibitem[{Youdin \& Lithwick(2007)}]{Youdin07}
Youdin, A.~N., \& Lithwick, Y. 2007, Icarus, 192, 588

\end{thebibliography}
\end{document}